\documentclass[twocolumn, apl, amsmath, amssymb, showpacs]{revtex4-2}
\usepackage{epsf}      
\usepackage{graphicx}
\usepackage{color}
\usepackage{soul}
\usepackage{gensymb}
\usepackage{sidecap}
\usepackage{amsmath}
\usepackage{mathtools}
\usepackage[colorlinks=true, linkcolor=blue, citecolor=blue, urlcolor=blue]{hyperref}
\begin{document}

\title{Structural, electronic, and magnetic properties of HoCo$_2$ thin films grown by pulsed laser deposition}
\author{Ajay Kumar}
\email{ajay1@ameslab.gov}
\affiliation{Ames National Laboratory, U.S. Department of Energy, Iowa State University, Ames, Iowa 50011, USA} 

\author{Christopher Burgio}
\affiliation{Ames National Laboratory, U.S. Department of Energy, Iowa State University, Ames, Iowa 50011, USA} 

\author{Anis Biswas}
\affiliation{Ames National Laboratory, U.S. Department of Energy, Iowa State University, Ames, Iowa 50011, USA} 

\author{Yaroslav Mudryk}
\affiliation{Ames National Laboratory, U.S. Department of Energy, Iowa State University, Ames, Iowa 50011, USA}

\date{\today}

\begin{abstract}

We report the growth and physical properties of HoCo$_2$ thin films on Si substrate by pulsed laser deposition. Well-defined Kiessig fringes in the x-ray reflectivity data yield a film thickness of $35 \pm 1$~nm and demonstrate smooth film surface and interface profiles. Atomic force microscopy confirms the high morphological quality of the films, revealing subnanometer surface roughness of $\sim$170~pm. Two independently prepared films exhibit a reproducible ferrimagnetic transition at $T_{\mathrm{C}}=77\pm3$~K, demonstrating retention of the characteristic magnetic ordering of bulk HoCo$_2$. However, the  absence of resolvable thermal hysteresis indicates a predominantly continuous, second-order like transition in the films, in contrast to the first-order transition of bulk HoCo$_2$. At 5~K, the films exhibit clear magnetic hysteresis with a coercive field of $\sim$600~Oe, substantially larger than the negligible coercivity of bulk HoCo$_2$. X-ray photoelectron spectroscopy reveals comparable Ho $4d$ and Co $2p$ core-level line shapes and broadly similar valence-band spectra for the bulk and film, indicating preservation of the essential local electronic environment and occupied electronic structure.

\end{abstract}

\maketitle

\section{\noindent ~Introduction}

The cubic Laves phases RCo$_2$ (R = rare earth) provide a well-established platform for studying the interplay among localized rare-earth magnetism, itinerant Co $3d$ electrons, and magnetoelastic coupling \cite{Gratz_JPCM_01, Khmelevskyi_JMCM_2000, Stein_JMS_21}. Across the series, the magnetic transition changes from predominantly second-order for lighter rare-earth members to first-order magnetostructural behavior for R = Dy, Ho, and Er \cite{Khmelevskyi_JMCM_2000, Inoue_JPFMP_88, Albillos_PRB_06, Kumar_JAP_24}. This first-order magnetic transition is accompanied by an exceptionally large magnetocaloric response, making the heavy rare-earth RCo$_2$ compounds among the most promising materials for cryogenic magnetic refrigeration applications \cite{Kumar_JAP_24, Albillos_PRB_06, Bykov_Jalcom_24}. The pronounced sensitivity of the transition to pressure, composition, and magnetic field makes the RCo$_2$ family particularly appealing for investigating how modifications of the electronic and structural environment control the order of magnetic phase transitions \cite{Syshchenko_JalCom_01, Forker_JPCM_06, Hauser_PRB_98}.


Extending rare earth--transition metal intermetallics to thin-film form offers additional routes to tune magnetic interactions through reduced dimensionality, strain, interfaces, and microstructure. In particular, strain can alter the thermodynamic nature of magnetic transitions. For example, in the rare earth double-perovskite Dy$_2$NiMnO$_6$, the first-order metamagnetic transition observed in the bulk transforms to second order in epitaxial films grown on SrTiO$_3$ and LaAlO$_3$ substrates, thereby suppressing the associated thermal-hysteresis loss during magnetocaloric cycling~\cite{Bhat_PRB_25}. In contrast, Gd$_5$(Si,Ge)$_4$ thin film retained the first-order magnetostructural transition of its bulk counterpart, while exhibiting a broadened magnetic response, an increased transition temperature, and markedly reduced magnetic hysteresis~\cite{Hadimani_APL_15}. Realizing these opportunities, however, requires careful control of film composition and morphology. Achieving a targeted composition in sputtered rare earth--transition metal films can be difficult owing to the sensitivity to deposition conditions~\cite{Inyang_APL_23, Wang_JSNM_21}. Moreover, the high fluence required to ablate the rare earth intermetallic targets using pulsed laser deposition can promote droplet formation and complicate the growth of smooth films~\cite{Shepelin_CSR_23}.

Recently, Skokowski \textit{et al.} reported PLD-grown Co-rich RCo$_2$ related Tb$_{31}$Co$_{69}$ and Dy$_{31}$Co$_{69}$ films on Si substrates~\cite{Skokowski_JMMM_26}.  Both films exhibited broad magnetic transitions at low temperatures compared with their bulk counterparts, particularly DyCo$_2$, which shows a first-order transition in bulk~\cite{Skokowski_JMMM_26, Khmelevskyi_JMCM_2000, Albillos_PRB_06}. These results illustrate the feasibility of producing RCo$_2$-related thin films, while highlighting the need for improved control of crystallinity, composition, and morphology to assess the intrinsic magnetic behavior of the Laves phase. In this context, HoCo$_2$ is a particularly suitable system, which undergoes a first-order ferrimagnetic transition near $T_{\mathrm{C}}\sim76$~K, accompanied by a cubic-to-tetragonal structural transformation \cite{Gignoux_PRB_75, Moon_JAP_65, Rubinstein_JMMM_81}. The transition can be driven toward second-order behavior by rare-earth substitution or hydrostatic pressure, highlighting its sensitivity to modifications of the Co $3d$ electronic states and magnetovolume coupling \cite{Syshchenko_JalCom_01, Forker_JPCM_06, Hauser_PRB_98}. Thin-film growth therefore provides a complementary approach to examine whether reduced dimensionality and film-specific structural constraints modify the spin-lattice coupling sufficiently to manipulate magnetostructural instability.

Here, we report the growth of HoCo$_2$ thin films on Si(100) substrate by pulsed laser deposition using a 248~nm KrF excimer laser. We combine x-ray reflectivity, atomic force microscopy, x-ray diffraction, SEM-EDS, magnetization, and x-ray photoelectron spectroscopy to establish the film quality, composition, magnetic behavior, and electronic structure. Particular attention is given to the retention of the characteristic $\sim77$~K transition and to its thermodynamic character in the thin-film geometry, both of which are essential for exploiting the giant magnetocaloric response of HoCo$_2$.

 \begin{figure*}
\includegraphics[width=6.8in]{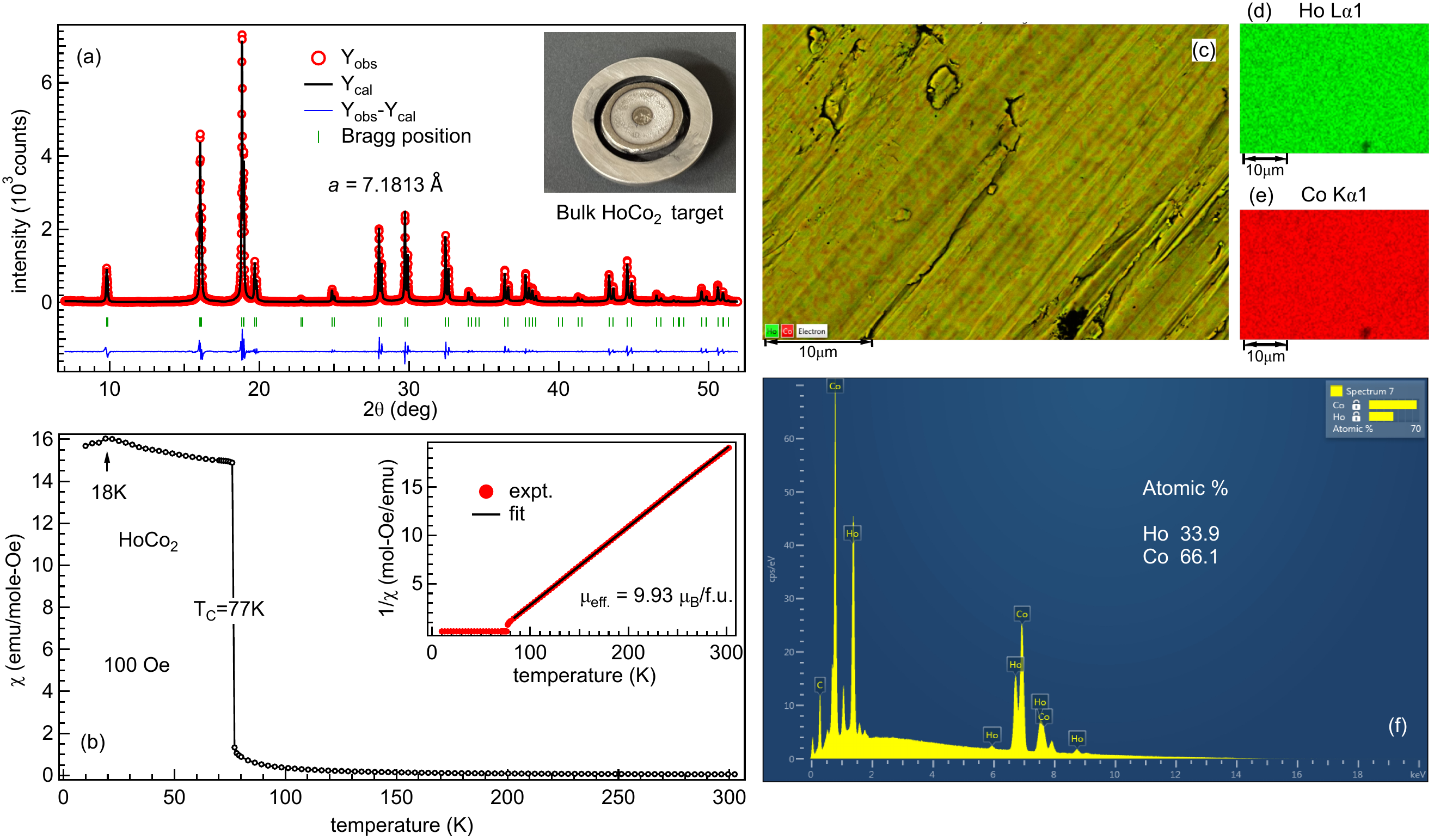} 
\caption {Structural, magnetic, and compositional characterization of the bulk HoCo$_2$ target used for thin-film deposition. (a)  Rietveld refinement of room temperature powder XRD pattern recorded using Mo $K_\alpha$ radiation. Inset shows the photograph of the bulk HoCo$_2$ target after PLD use. (b) Temperature dependent magnetic susceptibility measured at 100~Oe. Inset shows the inverse susceptibility with Curie-Weiss fit represented by the solid black line. (c) SEM-EDS elemental overlay map, (d, e) corresponding EDS elemental maps for  Ho and Co, respectively. (f) EDS spectrum of the target HoCo$_2$ sample. }
\label{Fig1}
\end{figure*}

\section{\noindent ~Experimental}

A bulk HoCo$_2$ sample of $\sim$25~g, used as the target for pulsed laser deposition (PLD), was synthesized by arc melting, as described in Refs.~\cite{Kumar_JAP_24, Mudryk_JMCC_16}. A circular PLD target of $\sim$ 20~mm diameter was prepared from this ingot by finely polishing both sides [see inset of Fig.~\ref{Fig1}(a)]. Prior to the first deposition, the target was pre-ablated to remove possible surface contamination. A naturally oxidized Si(100) substrate of dimensions $\sim$5 $\times$ 5 $\times$ 1~mm$^3$ was ultrasonically cleaned in acetone and methanol for 10~min each and then mounted on the sample holder using carbon paste. HoCo$_2$ films were deposited at a substrate temperature of 800~$^\circ$C under ultrahigh-vacuum conditions ($\sim$5 $\times$ 10$^{-9}$~Torr) using a 248~nm KrF excimer laser (Coherent). The deposition was performed using 50,000 laser pulses at a repetition rate of 10~Hz, with a fluence of $\sim$3~J/cm$^2$ and a target--substrate distance of $\sim$5.5~cm. After growth, the films were transferred and handled in an Ar-filled glovebox to minimize the air exposure. Two films were grown under identical deposition conditions to verify the reproducibility of the measured properties.

Powder x-ray diffraction (XRD) data for the target sample were collected using a Rigaku diffractometer with Mo $K_{\alpha}$ radiation~\cite{Holm_RSI_04}. The grazing incidence (GI) XRD and x-ray reflectivity (XRR) measurements on the thin-film samples were performed using a Rigaku SmartLab diffractometer equipped with Cu $K_{\alpha}$ source. The surface morphology and elemental distributions of the target and films were examined using a field-emission scanning electron microscope (FEI Teneo LoVac) equipped with Oxford energy-dispersive x-ray spectroscopy (EDS) and backscattered-electron detectors. Atomic force microscopy (AFM) measurements were performed using an Asylum Research Cypher ES Environmental AFM system. Temperature- and magnetic-field-dependent magnetization measurements were performed using a Quantum Design MPMS XL-7 superconducting quantum interference device (SQUID) magnetometer. The thin-film samples were mounted in plastic straws inside an Ar-filled glovebox, and the magnetic field was applied parallel to the film surface. X-ray photoelectron spectroscopy (XPS) measurements were performed using a Kratos Amicus/ESCA 3400 instrument with Al $K_{\alpha}$ radiation and a DuPont-type analyzer. The pass energy was set to 150~eV. The binding energy scale was calibrated by setting the C $1s$ peak to 284.8~eV. A Tougaard background was subtracted before fitting the core-level spectra. The asymmetric metallic Co $2p$ components were modeled using Doniach--\v{S}unji'c profiles, while symmetric multiplet and loss features were modeled using Voigt line shapes.

\section{\noindent ~Results and discussion}

 \begin{figure*}
\includegraphics[width=6.8in]{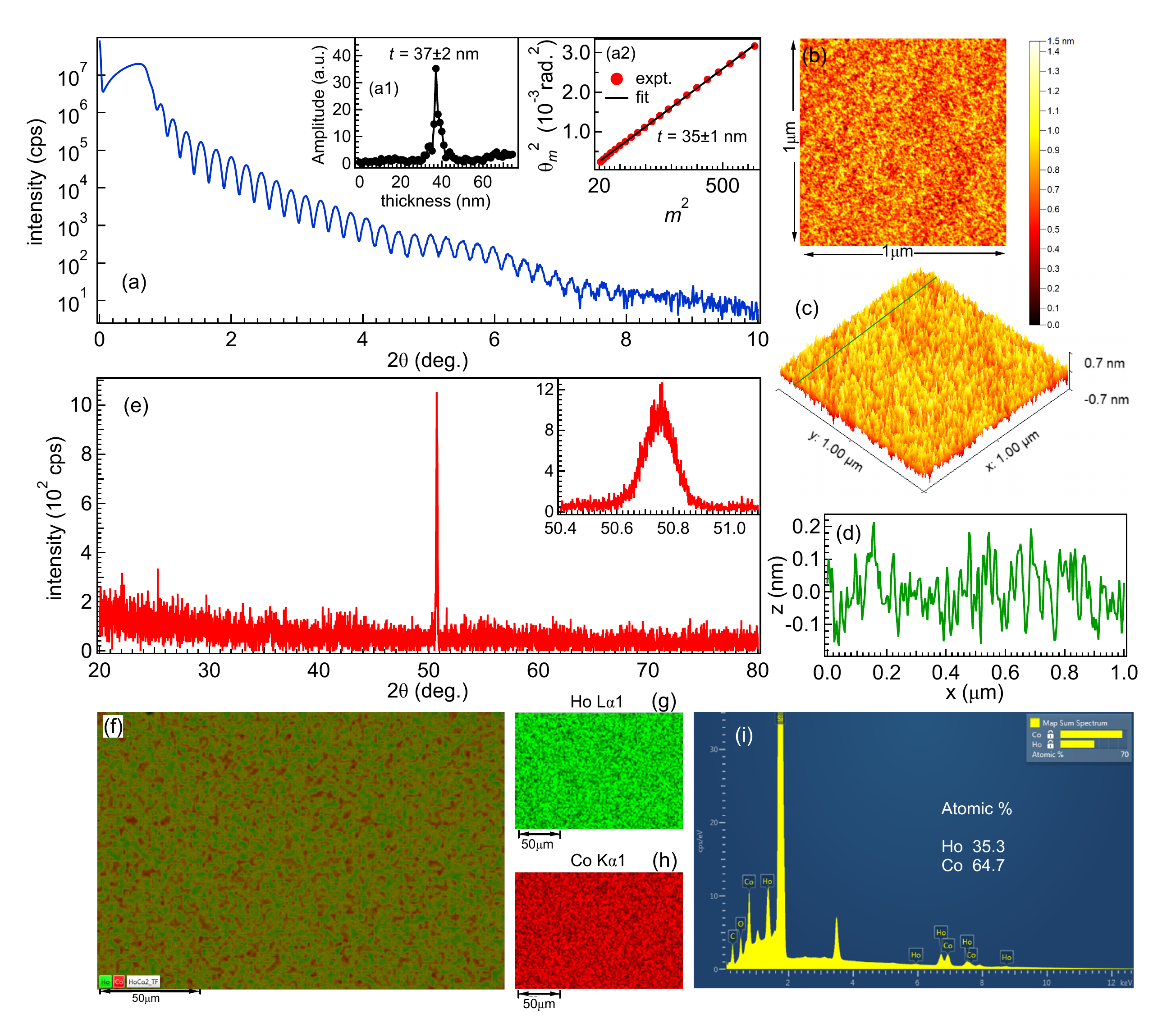} 
\caption {(a) X-ray reflectivity (XRR) pattern of the HoCo$_2$ film deposited on a Si substrate. Insets (a1) and (a2) show the fast Fourier transform of the XRR oscillations and the $\theta_m^2$ versus $m^2$ plot, respectively. The solid black line in (a2) represents the linear fit used to determine the film thickness. (b) Two-dimensional AFM topographic image of a $1 \times 1~\mu$m$^2$ area and (c) corresponding three-dimensional surface profile. (d) Height profile along the line marked in panel (c). (e) Room temperature GI-XRD pattern of the HoCo$_2$ thin film. The inset shows a slow scan of the reflection observed near $2\theta \approx 50.75^\circ$.  (f) SEM-EDS elemental overlay map and corresponding (g) Ho $L\alpha_1$ and (h) Co $K\alpha_1$ elemental maps. (i) EDS spectrum of the Si/HoCo$_2$ film. }
\label{Fig2}
\end{figure*}

The structural, magnetic and compositional properties of the bulk HoCo$_2$ target used to prepare the thin films are summarized in Fig. \ref{Fig1}. The Rietveld refinement of the room temperature XRD pattern recorded using Mo $K_\alpha$ radiation is presented in Fig. \ref{Fig1} (a), which confirms the single phase cubic MgCu$_2$-type C15 structure with space group $Fd\bar{3}m$, consistent with that reported in Refs. \cite{Gratz_JPCM_01, Mudryk_JMCC_16}. The refined lattice parameter is $a=7.1813$~\AA{}. The temperature dependent magnetic susceptibility, $\chi(T)$ measurements performed under an applied magnetic field of 100~Oe exhibits a sharp transition at $T_{\mathrm{C}}\approx77$~K [Fig.~\ref{Fig1}(b)], consistent with the reported first-order ferrimagnetic to paramagnetic transition in HoCo$_2$ \cite{Albillos_PRB_06, Mudryk_JMCC_16, Kumar_JAP_24, Gignoux_PRB_75}. A weaker low-temperature anomaly near 18~K, indicated by the arrow in Fig.~\ref{Fig1}(b), corresponds to the reported spin-reorientation transition \cite{Bykov_Jalcom_24, Mudryk_JMCC_16, Kumar_JAP_24}. The inverse susceptibility follows Curie--Weiss behavior between 80 and 300~K [inset of Fig. \ref{Fig1}(b)], yielding an effective magnetic moment, $\mu_{\mathrm{eff}}=9.93~\mu_{\mathrm{B}}$/f.u., reasonably close to the free-ion value of $10.61~\mu_{\mathrm{B}}$ for Ho$^{3+}$. The SEM image and the corresponding EDS elemental maps reveal a homogeneous spatial distribution of Ho and Co across the target sample [Figs.~\ref{Fig1}(c--e)]. The EDS spectrum yields atomic concentrations of 33.9\% Ho and 66.1\% Co [Fig.~\ref{Fig1}(f)], close to the expected 1:2 stoichiometric ratio for HoCo$_2$. Taken together, these results confirm that the bulk target possesses the expected crystal structure, characteristic magnetic behavior, and near-stoichiometric composition of HoCo$_2$.

 \begin{figure*}
\includegraphics[width=6.5in]{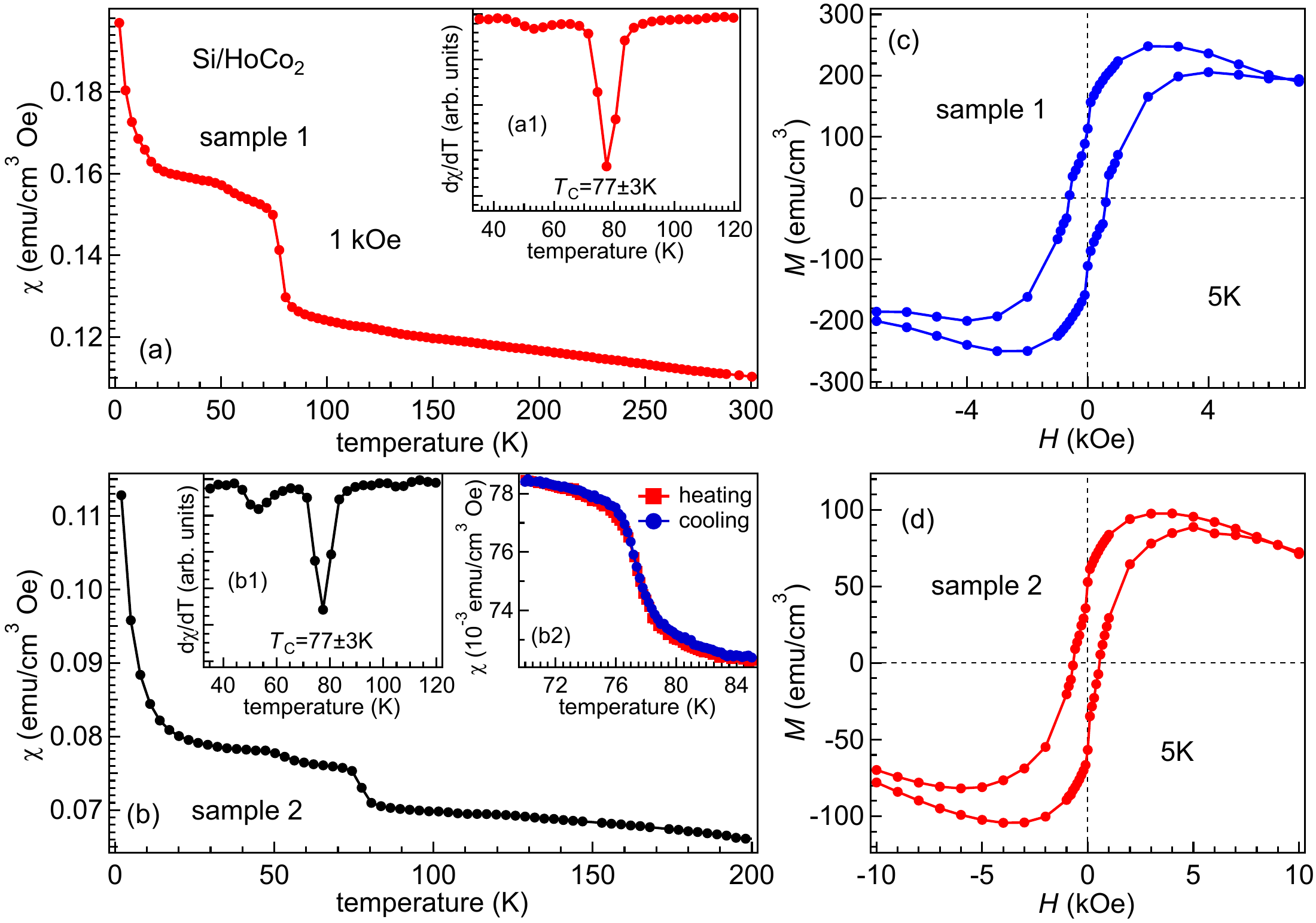} 
\caption {(a, b) Temperature dependent magnetic susceptibility, $\chi(T)$, measured under an applied field of 1~kOe for samples 1 and 2, respectively. Insets (a1) and (b1) show $d\chi/dT$ as a function of temperature for corresponding samples. Inset (b2) presents the heating and cooling $\chi(T)$ curves near $T_\mathrm{C}$. (c, d) Magnetic field dependent magnetization, $M(H)$, loops measured at 5~K for samples 1 and 2, respectively. }
\label{Fig3}
\end{figure*}

Figure~\ref{Fig2}(a) shows the x-ray reflectivity (XRR) pattern of the HoCo$_2$ thin film deposited on a Si substrate. Well-defined Kiessig fringes are observed over a broad angular range, indicating a continuous film with smooth surface and interface profiles \cite{Daillant_LNP_09}. In addition to the film-related fringes, the XRR pattern exhibits a broad modulation in the background, which may arise from the native SiO$_2$ layer on the Si substrate. The fast Fourier transform of the XRR oscillations yields a film thickness of $37\pm2$~nm [inset (a1)]. Furthermore, the film thickness was determined from the XRR oscillations using the modified Bragg equation \cite{Kushwaha_APL_17, Gupt_JVSTA_23}
\begin{equation}
\theta_m^2 = \theta_c^2 + \left(\frac{\lambda}{2t}\right)^2 m^2,
\end{equation}
where $m$ is the order of the XRR oscillation, $\lambda$ is the x-ray wavelength ($1.5406$~\AA{} for Cu K$_\alpha$ radiation), $t$ is the film thickness, $\theta_c$ is the critical angle, and $\theta_m$ is the angular position of the maximum of the $m^{\mathrm{th}}$ oscillation. All angles are expressed in radians. The $\theta_m^2$ versus $m^2$ plot for the HoCo$_2$ film is shown in inset (a2) of Fig.~\ref{Fig2}. A linear fit to the data in the range $m=5$--25, yields $t=35\pm1$~nm, which agrees well with that obtained from the FFT analysis.

The two- and three-dimensional AFM topographic images of the Si/HoCo$_2$ film, acquired over a $1\times1~\mu$m$^2$ area, are shown in Figs.~\ref{Fig2}(b, c), respectively. They reveal a very uniform and smooth surface with a root mean square (RMS) roughness of $\approx$170~pm. The height profile measured along the line marked in Fig.~\ref{Fig2}(c) exhibits only subnanometer scale height variations [Fig.~\ref{Fig2}(d)]. These results demonstrate the formation of a smooth and homogeneous HoCo$_2$ film, consistent with the well-defined Kiessig fringes observed in the XRR data.

The room temperature GI-XRD pattern recorded for the Si/HoCo$_2$ film at an incident angle of $0.1^\circ$ is shown in Fig.~\ref{Fig2}(e). A distinct reflection is observed at $2\theta \approx 50.75^\circ$, as shown more clearly in the slow scan in the inset of Fig.~\ref{Fig2}(e). This position corresponds to the $(400)$ Bragg reflection of the cubic HoCo$_2$ Laves phase, confirming a crystalline film contribution. Upon increasing the incident angle, this reflection weakens while Si-related reflections become more prominent, indicating an increased substrate contribution. As only one film-related reflection is resolved, the present data do not allow a full structural refinement or an unambiguous assessment of texture and phase purity. The SEM-EDS overlay map and the individual Ho and Co elemental maps demonstrate a homogeneous lateral distribution of both elements over the measured area [Figs.~\ref{Fig2}(f--h)]. The EDS spectrum yields approximate atomic concentrations of 35.3\% Ho and 64.7\% Co [Fig.~\ref{Fig2}(i)], indicating a slightly Ho-rich composition relative to the nominal 1:2 stoichiometry of HoCo$_2$. However, this small deviation may arise from the uncertainty of EDS quantification for a film of relatively low thickness, where the measured signal can be influenced by the underlying Si substrate.

The temperature-dependent magnetic susceptibility, $\chi(T)$, measured under an applied field of 1~kOe for two Si/HoCo$_2$ films (samples 1 and 2) deposited under identical conditions to ensure the reproducibility, is shown in Figs.~\ref{Fig3}(a, b). Both films exhibit a magnetic ordering transition near $T_{\mathrm{C}}\approx77$~K, which is more clearly identified from the $d\chi/dT$ curves shown in insets (a1) and (b1). The reproducible transition temperature confirms that the characteristic magnetic ordering of HoCo$_2$ is retained in films even with a thickness of $\sim$35~nm. In contrast, such a well-defined, bulk-like transition was not discernible in the Co-rich Tb$_{31}$Co$_{69}$ and Dy$_{31}$Co$_{69}$ films prepared by PLD, where only broad anomalies near $T_{\mathrm{C,Laves}}$ in $dM/dT$ were reported \cite{Skokowski_JMMM_26}. A weak but reproducible feature between 50 and 60~K observed in the present case [Figs. \ref{Fig3}(a, b)] may be associated with the freezing of residual oxygen in the SQUID chamber, which can become noticeable because of the small magnetic signal of the thin films. In addition, the sharp upturn below 10~K originates from paramagnetic impurities in the Si substrate. The data are presented without subtracting the substrate contribution to avoid uncertainties associated with the background correction. 

Interestingly, the heating and cooling $\chi(T)$ curves measured near $T_{\mathrm{C}}$ show no resolvable thermal hysteresis [inset (b2) of Fig.~\ref{Fig3}(b)]. This behavior is consistent with a predominantly continuous, second-order like magnetic transition in the thin films. In contrast, the bulk HoCo$_2$ sample exhibits a sharp transition with a clear thermal hysteresis of approximately 1~K \cite{Kumar_JAP_24}. These results suggest that reducing the dimensionality likely modifies the character of the magnetic transition relative to bulk HoCo$_2$.

The as-measured field dependent magnetization loops at 5~K for samples 1 and 2 are shown in Figs.~\ref{Fig3}(c, d), respectively. Both films exhibit clear magnetic hysteresis with a coercive field of $\sim$600~Oe, substantially larger than the negligible coercivity of bulk HoCo$_2$ \cite{Kumar_JAP_24}. This value is, however, close to the coercivities of $\sim$1000 and $\sim$900~Oe reported for Tb$_{31}$Co$_{69}$ and Dy$_{31}$Co$_{69}$ films at 150 and 80~K, respectively \cite{Skokowski_JMMM_26}. At higher fields, the diamagnetic contribution from the Si substrate becomes increasingly significant, as reflected in the high-field slope of the measured loops. Although the absolute magnetic signals differ between the two films, their transition temperatures and overall magnetic behavior are consistent, demonstrating the intrinsic nature of the observed magnetic response. The variation in signal magnitude may reflect small differences in the effective magnetic volume arising from variations in film thickness, composition, microstructure, interfacial or surface oxidation, and the substrate contribution.

\begin{figure*}
\includegraphics[width=6.8in]{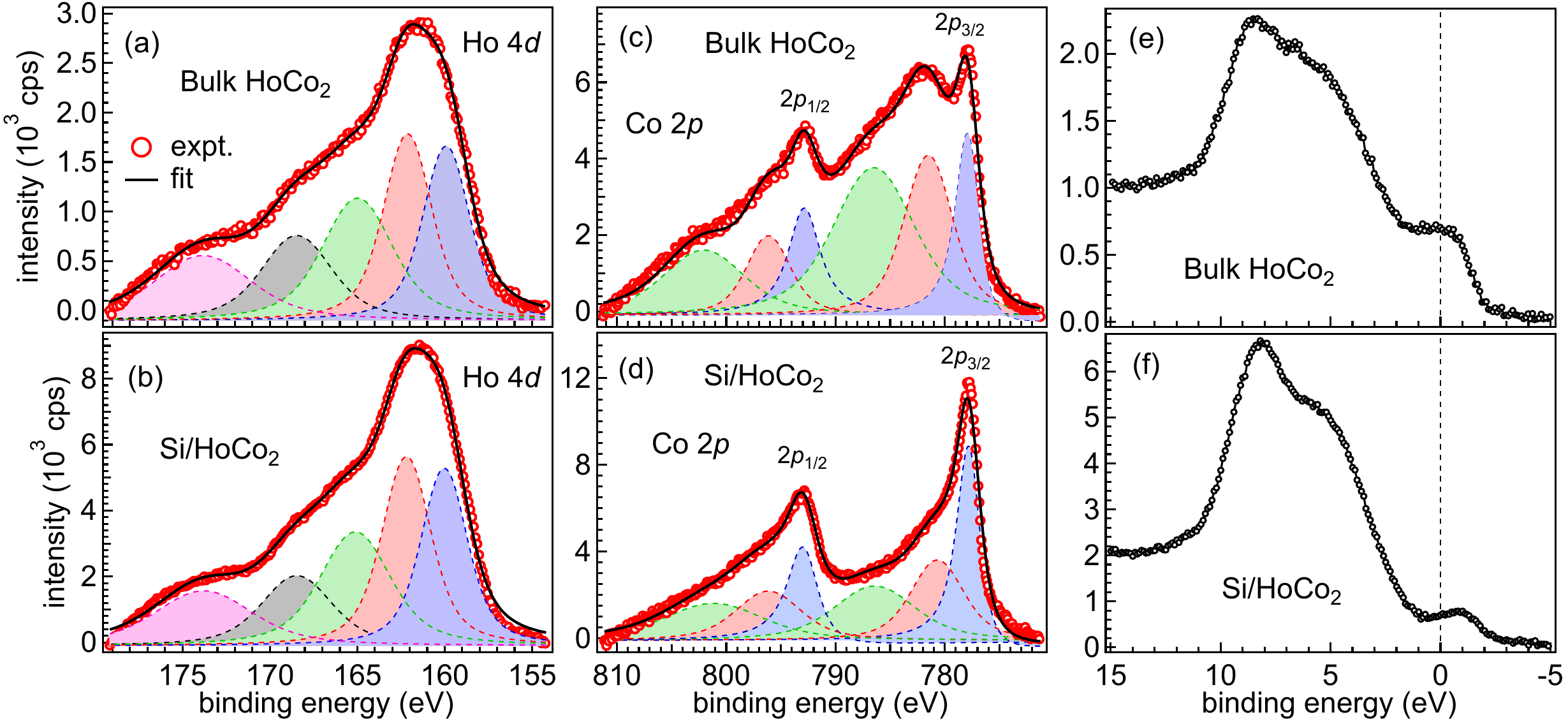} 
\caption {XPS spectra of the (a, b) Ho $4d$ and  (c, d) Co $2p$ core levels for bulk HoCo$_2$ and the Si/HoCo$_2$ film, respectively. Red open circles denote the experimental data, solid black lines represent the total fitted envelope, and shaded areas show the individual fitted components. (e, f) Valence band spectra of bulk HoCo$_2$ and the Si/HoCo$_2$ film, respectively.}
\label{Fig4}
\end{figure*}

The x-ray photoelectron spectroscopy (XPS) measurements were performed on both bulk HoCo$_2$ and the Si/HoCo$_2$ film to examine their electronic structure and surface chemical state. Figures~\ref{Fig4}(a, b) show the Ho $4d$ core-level spectra of bulk HoCo$_2$ and the Si/HoCo$_2$ film, respectively. Both spectra exhibit an intense, broad feature centered near 162~eV,  deconvolution of which resolves two relatively sharp components separated by $\sim$2.1~eV. In addition, lower intensity multiplet derived features are observed near 165.0 and 168.5~eV, together with a broader feature around 174.0~eV in both samples. Such a complex line shape is characteristic of strong multiplet interactions between the Ho $4d$ core hole and the partially filled Ho $4f$ shell, as reported for metallic Ho and Ho-based intermetallic compounds \cite{Chuang_PRB_20, Biswas_PRB_13, Morgan_SSS_24, Shishido_Jalcom_99}. The close similarity of the Ho $4d$ line shapes in the bulk and film indicates that Ho retains a comparable local electronic environment and trivalent state in the thin film. 

The Co $2p$ spectra of bulk HoCo$_2$ and the Si/HoCo$_2$ film [Figs.~\ref{Fig4}(c, d)] exhibit principal $2p_{3/2}$ and $2p_{1/2}$ features near 778 and 793~eV, respectively, with a spin-orbit separation of $\sim$15~eV. These binding energies are consistent with metallic Co-derived electronic states~\cite{Biesinger_ASS_11}. Both spin-orbit components display a clear asymmetric tail toward higher binding energy, which was modeled using Doniach--\v{S}unji'{c} line shapes \cite{Doniach_JPC_70}. In addition, weak higher binding energy loss features were modeled using symmetric Voigt profiles. Such asymmetric core level line shapes and plasmon loss features are characteristic of metallic Co~\cite{Biesinger_ASS_11, Grosvenor_IC_5}. The loss feature associated with the bulk $2p_{3/2}$ component is more pronounced than that observed for the film and the reported behavior in \cite{Grosvenor_IC_5}. Because the corresponding features are weak and partially overlap in the $2p_{1/2}$ region, a unique microscopic assignment of this intensity difference is not possible from the present spectra alone.

Figures~\ref{Fig4}(e, f) show the valence band spectra of bulk HoCo$_2$ and the Si/HoCo$_2$ film, respectively. Both spectra exhibit a broad valence band envelope with appreciable spectral weight near the Fermi level, consistent with the metallic character of HoCo$_2$. Two broad features are observed around 6 and 9~eV, which may contain contributions from hybridized Ho $4f$ and Co $3d$ states, together with more delocalized valence states. The broadly similar spectral shapes of the bulk and film suggest that the essential occupied electronic structure is retained in the film. Nevertheless, higher resolution ultraviolet photoelectron spectroscopy would be required to resolve the near Fermi level electronic structure more reliably.

\section{\noindent ~Conclusion}

In summary, HoCo$_2$ thin films with a thickness of $\sim$35~nm were grown on Si substrate by pulsed laser deposition. The well-defined Kiessig fringes in the x-ray reflectivity data and the subnanometer surface roughness measured by atomic force microscopy demonstrate the high morphological quality of the films. SEM-EDS reveals homogeneous Ho and Co distributions, with a composition of 35.3 at.\% Ho and 64.7 at.\% Co, indicating a slightly Ho-rich film relative to the nominal HoCo$_2$ stoichiometry. The thin films retain the ferrimagnetic transition near $T_{\mathrm{C}}=77\pm3$~K. However, the absence of resolvable thermal hysteresis indicates a predominantly continuous, second-order like transition, in contrast to the first-order magnetostructural transition of bulk HoCo$_2$. The films further exhibit an enhanced coercive field of $\sim$600~Oe at 5~K. XPS measurements reveal similar Ho $4d$ multiplet structure, metallic Co-derived core-level features, and broadly comparable valence-band spectra in the bulk and film, indicating that the essential electronic structure is retained after thin-film growth.

Future investigations of films with different thicknesses will be important for identifying a possible critical thickness for the first-to second-order like change in the transition character. Temperature dependent structural measurements across $T_{\mathrm{C}}$ will also be important for determining whether the cubic to tetragonal structural transformation accompanying the 77~K transition in bulk HoCo$_2$ persists, is modified, or is suppressed in the thin film geometry. Such studies will be essential for understanding how the giant magnetocaloric response of HoCo$_2$ evolves with reduced dimensionality.

\subsection{Acknowledgments}

This work was performed at Ames National Laboratory and was supported by the Division of Materials Science and Engineering of the Office of Basic Energy Sciences, Office of Science of the U.S. Department of Energy (DOE). Ames National Laboratory is operated for the U.S. DOE by Iowa State University of Science and Technology under Contract No. DE-AC02-07CH11358. 

\subsection{Data Availability}
The data that supports the findings of this study are available from the corresponding author upon reasonable request.

\end{document}